\documentclass[3p]{elsarticle}
\usepackage[utf8]{inputenc}
\usepackage[american]{babel}
\usepackage{amsmath}
\usepackage{amsfonts}
\usepackage{amssymb}
\usepackage{mathrsfs}
\usepackage{makeidx}
\usepackage{graphicx}  
\usepackage{xcolor} 
\usepackage{siunitx} 
\usepackage{float}
\usepackage{hyperref}
\hypersetup{
    colorlinks=true,
    linkcolor=blue
    }
\usepackage{url}
\usepackage{caption} 

\usepackage{array}
\usepackage{wasysym}
\usepackage{textcomp}
\usepackage{setspace}

\title{Exploration of trans-Neptunian objects using the Direct Fusion Drive}
\author[1,2]{Paolo Aime}
\ead{paolo.aime.95@gmail.com}
\author[1,2]{Marco Gajeri}
\ead{marco.gajeri@gmail.com}
\author[2,3,4]{Roman Ya. Kezerashvili}
\ead{RKezerashvili@citytech.cuny.edu}
\address[1]{Politecnico di Torino, Torino, Italy}
\address[2]{New York City College of Technology, The City University of New York, New York, USA}
\address[3]{The Graduate School and University Center, The City University of New York, New York, USA}
\address[4]{Samara National Research University, Samara, Russian Federation}

\begin{document}
\begin{abstract}
The Direct Fusion Drive (DFD) is a nuclear fusion engine that will provide thrust and electrical power for any spacecraft. It is a compact engine, based on the D -$^{3}$He aneutronic fusion reaction that uses the Princeton field reversed configuration for the plasma confinement and an odd parity rotating magnetic field as heating method to achieve nuclear fusion (Cohen et al., 2019), which will heat the deuterium, also used as propellant. \par
In this work we present possibilities to explore the solar system outer border using the DFD. The objective is to reach some trans-Neptunian object, such as the dwarf planets Makemake, Eris and Haumea in less than 10 years with a payload mass of at least of 1500 kg, so that it would enable all kind of missions, from scientific observation to in-situ operations. For each mission a thrust-coast-thrust profile is considered. For this reason, each mission is divided into 3 phases: i. the spiral trajectory to escape Earth gravity; ii. the interplanetary travel, from the exit of Earth sphere of influence to the end of the coasting phase; iii. maneuvers to rendezvous with the dwarf planet. Propellant mass consumption, initial and final masses, velocities and $\Delta V$ for each maneuver are presented. Calculations to reach a vicinity at 125 AU for the study of Sun magnetosphere as well as Eris via flyby are also presented, with interest on the influence of different acceleration phases. 

Our calculations show that a spacecraft propelled by DFD will open unprecedented possibilities to explore the border of the solar system, in a limited amount of time and with a very high payload to propellant masses ratio.
\end{abstract}
\maketitle
\section{Introduction} 
Since the early days of space exploration, we have always tried to achieve destinations never reached before. If the next big destination for human exploration is Mars, on the robotic side of space travel we managed to achieve objectives that are further away. The most recent achievement is New Horizons mission, which visited Pluto at the inner border of the Kuiper Belt \cite{stern2018new, stern2015pluto,young2008new}. Launched about three decades before it, Voyager 1 and Voyager 2 are currently at about 148 AU and 124 AU from the Sun, respectively, and have gathered a tremendous amount of scientific data \cite{voyager1nasa,voyager2nasa}. A mention should also go to Pioneer 10 and 11 missions, which also traveled through the Kuiper Belt, even though the communication with those spacecrafts was becoming very feeble. The last signals received from Pioneer 10 and 11 are dated back to 2003 at 84 AU and 1995 at 45 AU, respectively \cite{doody2010deep}.\par

The missions mentioned above relied on chemical propulsion systems. The main features of a mission planned with such technology are an initial, enormous velocity gain combined with one or more velocity gain from planets flybys. The first is very demanding in terms of propellant mass at launch, and often it is not acceptable for ambitious missions. Electric propulsion systems show a huge improvement in terms of  capabilities, because they enable a small, but continuous, acceleration to the spacecraft which can theoretically achieve competitive velocities for long distance travel. It has already been used for major missions such as BepiColombo \cite{benkhoff2010bepicolombo} and Dawn \cite{russell2011dawn}. Building extremely powerful electric thrusters has no relevant physical limitations, the only insuperable impediment at present is the power source to feed them. Electric propulsion systems nowadays are suitable only for exploration of the inner solar system because they can be easily fed by a generation system relying on sunlight, which is a potentially infinite power source. Further away than Jupiter orbit, a solar-powered electric thruster is no longer a feasible option, because solar arrays would grow exponentially in size. For this reason, the only huge breakthrough in space propulsion would be a nuclear power based system. In these terms, we could either think of a nuclear power source that is used to feed an electric thruster, a nuclear electric propulsion system (NEP), or we can think of directly converting the heat from the reaction to thrust with a nuclear thermal propulsion system (NTP). \par

In this work we present a new class of missions that will be enabled by the realization of the Direct Fusion Drive (DFD). This device, under research at Princeton Plasma Physics Laboratory (PPPL) and Princeton Satellite Systems (PSS), is an extremely promising nuclear fusion reactor that will generate thrust as well as electrical power for the spacecraft. Furthermore, it will be extremely compact, compared to other nuclear fusion reactors \cite{cohen2017direct,thomas2017fusion}. Its main feature is that thrust is achieved via a direct process, in which the propellant fluid is heated by the fusion products and then directly expanded into the nozzle. The latter, all together, will grant thrust to the spacecraft. This system should be considered halfway between a NTP and an electromagnetic thruster, because the propellant is heated via an exchange mechanism, but the thrust generation process is mainly due to electromagnetic phenomena. \par

The paper is organized in the following way: in Section \ref{sec:just} are given scientific justifications for missions to the recently discovered trans-Neptunian objects - dwarf planets Haumea, Makemake and Eris and an exploration of 125 AU destination and beyond. A brief description of a DFD concept and a thruster’s main characteristics are presented in Section \ref{sec:DFD}. In Section \ref{sec:TNOs} and \ref{sec:125AU} we consider missions design for trans-Neptunian planets and 125 AU destinations, respectively, and results of calculations and analyses of thrust-coast-thrust profiles for each mission are presented. Finally, conclusions follow in Section \ref{sec:conc}.
\section{Scientific justification} \label{sec:just}
\subsection{Trans-Neptunian objects}
A class of bodies that has never been explored, except for the New Horizons mission, is the one of trans-Neptunian objects (TNOs), which is composed of those asteroids and dwarf planets whose orbits extend from the orbit of Neptune towards the interstellar space, that means from 30 AU on. Of those bodies, many have their orbit entirely within the Kuiper Belt, which extends from about 30 to 50 AU \cite{nasakuiper2019,delsanti2006solar}. Depending on the orbital resonance with Neptune, their inclination and eccentricity may vary \cite{delsanti2006solar} and they can be classified in different families depending on those characteristics. Among the most interesting objects, beyond the well known Pluto, there are the well identified dwarf planets Haumea, Makemake, Eris, Quaoar and Sedna. In this work we focused on the three biggest dwarf planets, which are Haumea, Makemake and Eris. Those have all been discovered and studied only from the early years of the new millennium  \cite{brown2005discovery, delsanti2006solar, rabinowitz2006photometric, ortiz2012albedo, nesvorny2016orbital,  nasaeris2019, nasahaumea2019, nasamakemake2019}.  In Table \ref{tab:dwarf_planets} main physical and orbital parameters are reported for the three subject dwarf planets and compared to those of Pluto. All TNOs ephemeris are taken from Ref. \cite{JPLHorizons}.  \par

\begin{table} [h]
\centering
\caption{Main physical and orbital properties for objective dwarf planets \cite{brown2005discovery, delsanti2006solar, rabinowitz2006photometric, ortiz2012albedo, JPLHorizons, nasaeris2019, nasahaumea2019, nasamakemake2019}. When the dwarf planet has a near spherical shape, its diameter is reported, as in the case of Pluto and Eris. Haumea is an ellipsoid, and for this reason the three axis are reported. Makemake is a sphere flattened at the poles. }
\begin{tabular}{r c c c c}
\hline \hline
 & Pluto & Haumea & Makemake & Eris \\ 
\hline  
Mass, [kg] & $1.30\times 10^{22}$ & $4.01\times 10^{21}$ & $3.1 \times 10^{21}$ & $1.66 \times 10^{22}$ \\
Dimensions, [km] & $\diameter = 2376  $& $ 2100\times 1680 \times 1074$& $1502 \times 1430$ & $\diameter = 2326$ \\ 
Semi-major axis [AU] & 39.482 & 43.287 & 45.561 & 67.740 \\
Aphelion, [AU] & 49.305 & 51.600 & 52.761 & 97.468 \\ 
Perihelion, [AU] & 29.658 & 34.973 & 38.360 & 38.013 \\ 
Eccentricity & 0.249 & 0.192 & 0.158 & 0.439 \\ 
Inclination, [deg] & \ang{17.16} & \ang{28.12} & \ang{28.98} & \ang{44.14} \\ 
\hline \hline
\end{tabular} 
\label{tab:dwarf_planets}
\end{table}
Besides their dimensions and masses, those icy worlds are interesting because they could be hosting subsurface oceans \cite{hussmann2006subsurface}. Furthermore, the possibility to study those dwarf planets for many years by following them during their path around the Sun could reveal important information about the origins of the solar system, its evolution, as well as about the nature of the Kuiper Belt itself. Throughout the years, very few missions have been studied to reach those worlds, mostly with classical propulsion systems \cite{dankanich2010electric, mcgranaghan2011survey, gleaves2012survey, kreitzman2013mission, baskaran2014survey}.
\subsection{125 AU destination}
Another important scientific objective beyond the orbit of Neptune is the study of the external border of the heliosphere,  which in some way can be considered as an interstellar objective. It is the region of space that is formed by the solar wind, which is formally plasma, which surrounds the Sun. This region is basically a bubble inside the interstellar medium (ISM), composed of the matter and the radiation that are present in the space between different star systems \cite{cox1987local, frisch1995characteristics}. The amount of galactic cosmic rays inside the heliosphere is milder than outside because it works as a shield from heavier radiation from distant stars.  \par

The magnetosphere follows the Sun in its movement across the galaxy, so its shape is not properly a sphere, it is more a comet-like shape. We can identify three different regions, all associated with the interaction of the solar wind plasma with the interstellar medium, which is also a plasma, but with smaller densities. This leads to magneto fluid dynamic phenomena such as shock waves and more. Those phenomena have been studied for years, both from Earth, with the interstellar boundary explorer mission (IBEX) \cite{mccomas2009ibex}, and on-site by the two Voyager missions. Those far space regions will also be visited in the near term future by the New Horizons spacecraft \cite{stern2018new}. The three main regions are: 
\begin{enumerate}
\item\textbf{\textit{The termination shock.}} The pressure difference between the interstellar medium and the solar wind plasma causes a shock wave. This zone is expected to lie from 75 to 90 AU from the Sun and, at present, only Voyager 1 and Voyager 2 spacecraft crossed it \cite{mccomas2019termination}.
\item\textbf{\textit{The heliosheath.}} This is the zone beyond the termination shock in which the solar wind interacts with the interstellar medium. It is likely formed by many bubbles of about 1 AU of width. These are probably caused by the fact that the Sun spins and its magnetic field rotates with it, and it becomes twisted and wrinkled at huge distances \cite{edgebigsurprire2011, edgemagbubble2011, burlaga2018heliosheath}. The heliosheath is considered to lay at 80-100 AU in the closest point.
\item\textbf{\textit{The heliopause.}} This is the zone where the solar wind is stopped by the ISM and pressures become equal. After this ``barrier," the amount of charged particles from solar wind decreases very steeply and the amount of galactic cosmic rays increases. Voyager 1 seems to have crossed this boundary in 2012 at a distance of 121 AU \cite{richardson2019voyager, burlaga2019magnetic}. 
\end{enumerate}
All of the previous regions are a very important subject of study to develop a better knowledge on how the solar wind and the Sun magnetosphere behave near the border with the ISM. Many other data can be obtained from a mission of this kind \cite{gruntman2006innovative}. Multiple missions to these destinations have been proposed, both with typical and innovative propulsion systems \cite{nock1987tau, liewer2000nasa, wallace2000interstellar, mcnutt2003realistic, lyngvi2004interstellar, mcnutt2005innovative,  fichtner2006science,  mcnutt2006innovative, ancona2019extrasolar}. \par
\section{Direct Fusion Drive} \label{sec:DFD}
The DFD is a revolutionary steady state nuclear fusion propulsion concept, based on the D - ${}^3$He aneutronic nuclear fusion reaction. The first studies related to a DFD can be traced back to Chapman's work  \cite{chapman1989fusion}. Princeton Plasma Physics Laboratory is developing its own version of DFD, with many improvements since Chapman's work \cite{chapman1989fusion}. In their configuration, the DFD will provide thrust of the order of $10^0-10^1$ N with specific impulses between $10^3-10^5$ s, as well as auxiliary power to the spacecraft systems, needed for station keeping and communications.  \cite{cohen2017direct}. \par
 
This engine is linear and the propellant is deuterium. In this DFD the propellant is first ionized, then it enters a region with a strong, externally-imposed, magnetic field. Here, the propellant flows around the engine's core, inside which nuclear fusion reaction occurs and its products heat up the propellant. Then, the hot propellant expands into a magnetic nozzle, producing thrust \cite{razin2014direct, cohen2017direct}. \par

This DFD is based on the Princeton field reversed configuration (PFRC), which is a field reversed configuration (FRC), where the plasma is heated by rotating magnetic fields with odd parity (RMF$_{\text{O}}$) \cite{cohen2000ion}.
A FRC is a toroidal plasma configuration, where the poloidal magnetic field is of many orders of magnitude higher than the toroidal one, and for this reason the latter becomes negligible \cite{romanelli2005assessment}. This configuration allows the plasma to achieve high $\beta$, which is defined as the ratio of plasma pressure to magnetic field energy density. High $\beta$ allows to reach high specific powers with the aneutronic D$- {}^3$He fusion reaction.

The FRCs rely on strong toroidal current that induces a closed poloidal field around the torus, that confines the plasma. All the plasma is contained into an open magnetic field (OMF) that is generated by external superconductive coils. In this way there is a region inside the OMF with a closed magnetic field. The innovation in the PFRC is that antennas that generate a rotating magnetic field (RMF) are placed around the plasma: their role is indeed to generate a radial RMF that excites the plasma and let it achieve nuclear fusion-relevant temperatures \cite{jones1999review, cohen2000ion, cohen2000maintaining}. This RMF would have odd parity with respect to a plane perpendicular to the axis: antennas are organized in an 8-shape, where two windings carry the current in opposite directions. The odd parity rotating magnetic field provides stability to the plasma, thus increasing its confinement time \cite{cohen2000maintaining,guo2005observations}. The latter is an important factor for nuclear fusion. It is important to point out that antennas are fixed in the space, and it is the magnetic field which is rotating, achieved by phasing different antennas of \ang{90} around the plasma region. \par

Let us consider the reactions admitted in D--$^{3}$He plasma. The
main reaction that happens inside the toroidal D--$^{3}$He plasma is the aneutronic fusion reaction 
\begin{eqnarray}
\text{D}+\text{}^{3}\text{He }=\text{ }^{4}\text{He (3.52 MeV)}+p\text{
(14.7 MeV),}  \label{D3He}
\end{eqnarray}
which does not produce any neutron by itself. This makes this reaction aneutronic, even though some neutrons are produced by side reactions 
\begin{eqnarray}
\text{D}+\text{D} &=&{ }^{3}\text{He (0.82 MeV)}+n\text{ (2.45 MeV),}
\label{DD3He} \\
\text{D}+\text{D} &=&\text{ }\text{ }\text{ T (1.01 MeV)}+p\text{ (3.03 MeV)}.  \label{DDT}
\end{eqnarray}%
One should mention that reaction (\ref{DDT}) involves the following secondary processes in the D--$^{3}$He plasma: 
\begin{eqnarray}
\text{D}+\text{T} &=&\text{ }^{4}\text{He (3.52 MeV)}+n\text{ (14.7 MeV),}
\label{DT} \\
\text{T}+\text{T} &=&\text{ }^{4}\text{He (3.52 MeV)} + 2n,  \label{TT}
\end{eqnarray}%
which also produce undesired neutrons. In Eqs.~(\ref{D3He}) - (\ref{TT}) values in parenthesis are the energy
of that particular fusion product. Fast neutrons carry off a significant part of released energy as well. Therefore, the undesired tritium produced via reaction (3) will increase the neutron flux due to reactions (\ref{DT}) and (\ref{TT}), which are secondary for D--$^{3}$He plasma. In Refs. \cite{khvesyuk1995ash,sawan2002impact} a method of tritium removal from the
plasma before it can fuse has been proposed.
Fortunately, those neutrons in process (%
\ref{DD3He}) are not very energetic, but still some
shielding is required, and enriched B$_{4}$C boron carbide seems to be a
good and affordable solution \cite{ThomasPSS2017,cohen2015reducing}. In order to further reduce the
probability of side reactions (\ref{DD3He}) and (\ref{DDT}) with neutrons,
the ratio of $^{3}$He to D is chosen to be 3:1.
The main product of the fusion would be highly energetic protons, that can be easily controlled with magnetic fields. Highly energetic particles, that have higher radius, are responsible for propellant heating in the scrape-off layer (SOL). This is the zone right outside fusion plasma where the cold propellant flows, and it is some centimeters thick. In that region propellant electrons are heated up by highly energetic fusion products. Those electrons exchange energy with ions starting from the nozzle throat  \cite{cohen2017direct,ThomasPSS2017,mcgreivy2016}, which will then convert it to kinetic energy during the expansion. This process is called thrust augmentation and involves roughly 50\% of power output from fusion.  \par

The main concern about the DFD is linked to availability of $^{3}$He on the Earth, which is very limited \cite{kennedy2018interstellar}, but it seems that new sources of it could be found on the Moon \cite{wittenberg1986lunar}. 

Many studies and simulations have been performed to estimate DFD performances at PSS and PPPL \cite{ThomasPSS2017,mcgreivy2016,thomas2017fusion,thomas2018nuclear}, mainly with UEDGE software \cite{osti_15007243,rognlien1992UEDGE}. Exact values for thrust and specific impulse depend on the percentage of fusion power which transfers to SOL. The lowest estimated power for this engine is the 1 MW class, while the highest possible is likely to be 10 MW.  \par

\begin{table} [h]
\caption{DFD characteristics for lowest and highest power configurations. Higher specific impulse leads to lower thrust for the same configuration \cite{cohen2017direct}.}
\centering
\begin{tabular}{r c c c c }
\hline \hline
 & \multicolumn{2}{c }{Low power} & \multicolumn{2}{c}{High power} \\
\hline 
Fusion power, [MW] & \multicolumn{2}{c}{1} & \multicolumn{2}{c }{10} \\ 
Specific impulse, [s] & 8500 & 8000 & 12000 & 9900 \\ 
Thrust, [N] & 4 & 5 & 35 & 55 \\  
Thrust power, [MW] & \multicolumn{2}{c}{0.46} & \multicolumn{2}{c}{5.6} \\ 
Specific power, [kW/kg] & \multicolumn{2}{c}{0.75} & \multicolumn{2}{c}{1.25} \\
\hline \hline
\end{tabular} 
\label{tab:DFDcharacteristics}
\end{table}

\section{Mission Design for TNOs destinations} \label{sec:TNOs}
\subsection{Trajectory design}
For this study let us select an engine of 2 MW of fusion and assume that thrust and specific impulse are constant. The latter assumption significantly simplify the trajectory design. The power transferred to the SOL is supposed of about 1 MW, with the resulting performance of 8 N of thrust and approximately 10000 s of specific impulse.  \par

The trajectory profile chosen is the simplest possible, the so called thrust-coast-thrust profile. This means that the engine produces thrust only for the first and last phases of the mission. For our purposes, it is not really important whether the engine is completely off or if it only does not use the propellant.  Let us divide the mission into three phases and study the $\Delta V$, fuel consumption and duration for each phase. Note that this division has only the purpose of better understanding the propulsion requirements in each part of the mission. The phases are the following:
\begin{figure} [h]
\centering
\includegraphics[width=\linewidth]{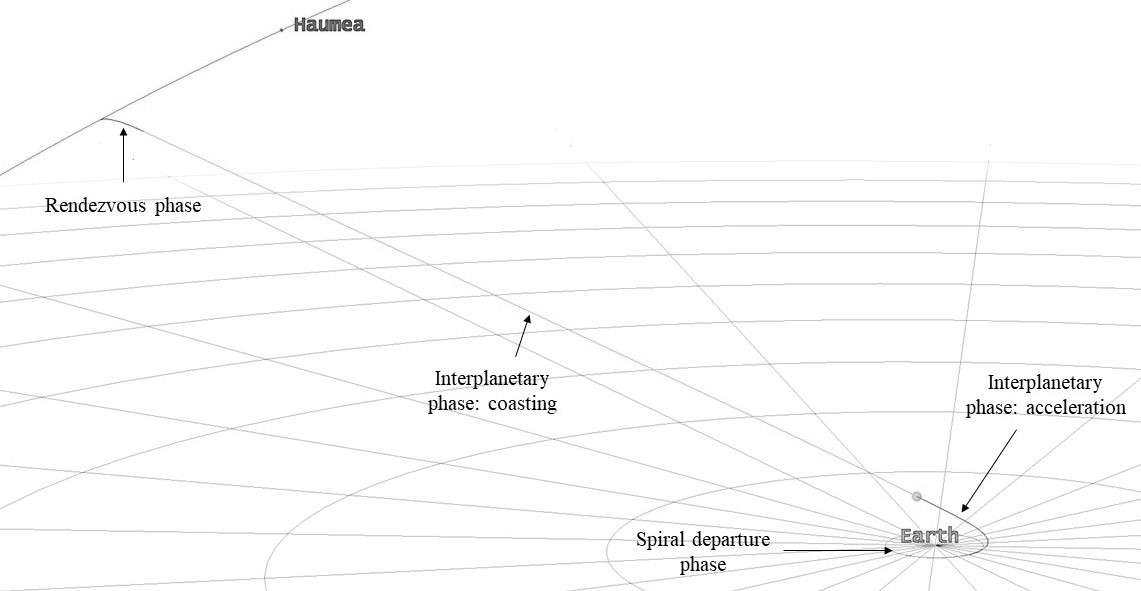}
\caption{The trajectory to reach Haumea. Each phase is here shown. The Earth orbit and a section of Haumea orbit are visible in this figure, respectively in the bottom right and top left. The visible grid represents the ecliptic plane. }
\label{fig:traj}
\end{figure}
\begin{enumerate}
\item\textbf{\textit{Spiral departure phase.}} This phase begins with the departure from a circular low Earth orbit (LEO) and ends with the exit from the sphere of influence of the Earth, which is supposed to be at about $10^6$ km from the center of the Earth. We consider that moment as the passage from the geocentric reference system to the heliocentric reference system. In this phase,  the thrust vector is always parallel to that of velocity. Even though it is not an optimal solution, it is a reasonable approximation of it \cite{boltz1992orbital}. The spacecraft velocity vector at the end of this maneuver is entirely contained into the ecliptic plane and also it is parallel to that of the Earth in a heliocentric reference system to maximize the velocity. All the perturbations in this phase are neglected and only the gravitational influence of the Earth is considered.
\item\textbf{\textit{Interplanetary phase.}} This is the continuation of the spiral departure phase and there is no real interruption in the DFD thrust. The further acceleration has both the purpose of bringing the spacecraft to high velocities as well as achieving the required inclination, by pointing the thrust vector slightly outside the ecliptic plane. After this acceleration, it follows a coasting phase, during which the spacecraft travels for many years towards its destination, in a hyperbolic orbit with respect to the Sun.
\item\textbf{\textit{Rendezvous phase.}} This phase consists of all the maneuvers necessary to rendezvous with the dwarf planet destination. First of all, we intent the word ``rendezvous" in the terms of a spacecraft that is be placed in an orbit extremely close to that of the object to study with respect to the Sun. This means that any maneuver to reduce the spacecraft orbit around the destination is not here analyzed. The main objective for this phase is to slow down the spacecraft and turn its trajectory to match that of the object. The optimal maneuver would have a thrust direction fixed in time for the entire maneuver, with both a radial and an anti-velocity component. However, the numerical solution could not be found in this way mainly due to the enormous velocities considered, which make the search for an exact value for thrust direction very difficult to converge. As an alternative, a proper steering law could be developed, but it is out of the scope of this article, and it will be addressed in future work. Instead we approximated this maneuver as a first pure deceleration phase, where the thrust is pointed against the velocity vector, followed by a further deceleration, where the thrust has a radial component that enables the trajectory to turn in time. In this case, thrust direction is inertially fixed. After this phase, the spacecraft is supposed to follow the destination dwarf planet for several years in its orbit around the Sun.
\end{enumerate}

\subsection{Inputs and results for TNOs missions}
The inputs for all calculations are the engine performance and the launch mass. The objective is to bring at least 1500 kg of payload at destination in less than 11 years. It is worth noticing that all trajectory calculations are performed on Systems Tool Kit (STK) Astrogator by AGI. The 2 MW DFD, under the hypothesis of a specific power of 1 kW/kg, results in approximately 2000 kg of engine, thereby a total dry mass of 3500 kg. At this point, the first estimation of propellant needed is 3988 kg, for a total launch mass of 7488 kg. The initial propellant mass estimate is achieved through an iterative process based on the solution of the Lambert problem. First, the overall $\Delta V$ is calculated, given the time of flight. After that, the propellant mass is calculated through the specific impulse, assuming the total days of thrust. After many iterations, we define an initial guess for the total thrust days and the propellant mass. This preliminary study enables us to understand that, given the time and payload masses constraints of our missions, a trajectory with continuous thrust is not achievable with the 2 MW class engine. In particular, we would need a higher specific impulse, otherwise the propellant mass at launch would be so high that the escape from LEO would become too challenging. This is a representative first approximation only because the acceleration and deceleration times are a minor part of the overall flight time. In order to simplify calculations, we always use the initial estimation of propellant mass for all the missions, and at the end of the design phase all of the unused propellant mass is allocated as payload mass. This means that the initial guess for the payload mass can only increase, if the propellant mass is enough to reach the destination. \par

For each mission, the three phases are studied. Results for the first phase are the same for all the missions and they are summarized in Table \ref{tab:spiral}, where we also present the time spent in the inner Van Allen belt, that is one of the most dangerous segments of the mission in terms of radiations and disturbance. The only difference between the three missions is the departure date. This date is chosen to be later than the year 2050 because the three planets are moving towards the line of nodes, and the later will be the departure, the less inclination change will be required. Though, this is not a big restriction, because the difference is only of some days of acceleration, and for this reason it is negligible.  \par

\begin{table} [h]
\centering
\caption{Maneuver duration, $\Delta V$ and propellant consumption for the phase of the spiral trajectories. It is also reported the calculated time inside the most dangerous part of the inner Van Allen belt. The boundaries of the most dangerous zone of the inner Van Allen belt are considered to be 1000 km and 6000 km.}

\begin{tabular}{rc}
\hline \hline
& Spiral phase \\
\hline
Duration & 79 days \\
$\Delta V$ & 7.54 km/s \\
Propellant consumption & 554 kg \\
Van Allen belt time & 18 days \\
\hline \hline
\end{tabular}
\label{tab:spiral}
\end{table}

\begin{figure} [h]
\centering
\includegraphics[width=1\linewidth]{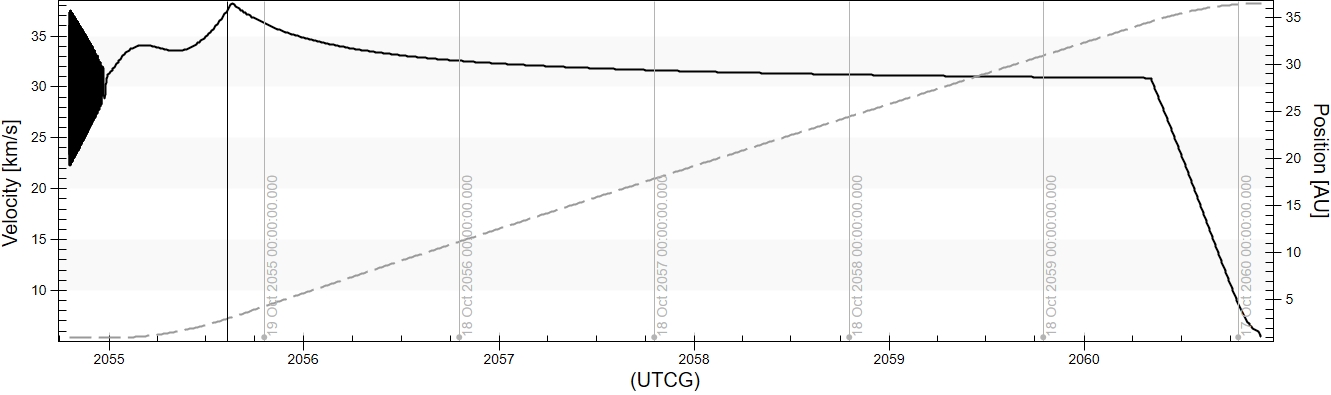}
\caption{Velocity (continuous line) and distance (dashed line) with respect to the Sun for Haumea mission. The initial black triangle is the result from the orbital motion around the Earth.}
\label{fig:haumea28inc_vel_pos}
\end{figure}

\begin{table} [h]
\caption{Summary for the three real case missions, with focus on the interplanetary and rendezvous phase maneuvers. It is worth noticing that the three sets of results are for scenarios with the same initial mass of 7488 kg. The final velocity for the interplanetary phase is the velocity at the end of the acceleration. The final coasting velocity is the one after the gravity losses, before rendezvous maneuvers. The final rendezvous velocity is the one at the end of the rendezvous phase. In the mission summary the time of flight is the total flight duration from the departure from LEO, the inclination is referred to the dwarf planet with respect to the ecliptic plane, the final mass delivered is the payload mass plus both the engine mass and the propellant mass not used. The $\Delta V$ in the mission summary is the sum of that of all thrust phases, including the spiral phase given in Table \ref{tab:spiral}.}
\centering
\begin{tabular}{lrccc}
\hline \hline
 &   & Haumea & Makemake & Eris \\ 
\hline 
Interplanetary  & Acceleration duration, [days] & 225  & 251 & 279 \\ 
phase & Propellant consumption, [kg] & 1586 & 1769 & 1966 \\ 
 & $\Delta V$, [km/s] & 25.47 & 28.89 & 32.71 \\ 
 & Final velocity, [km/s] & 38.22 & 37.86 & 43.33\\
 & [AU/year] & 8.05 & 7.98 & 9.13 \\
 & Coasting duration, [years] & 4.75  & 5.83 & 8.67 \\
\hline 
Rendezvous  & Maneuver duration, [days] & 206 & 204 & 243 \\ 
phase & Propellant consumption, [kg]  & 1455 & 1436 & 1709  \\ 
 & $\Delta V$, [km/s] & 31.15  & 31.96  &  41.37\\ 
 & Final coasting velocity, [km/s] & 30.87 & 30.99 & 38.52\\
 & Velocity at rendezvous, [km/s] & 5.35 & 4.58 & 3.09\\
\hline 
Mission  & Time of flight, [years] & 6.08 & 7.25 & 10.33 \\
summary  & Distance at rendezvous, [AU] & 36.51 & 44.35& 78.20\\ 
 & Inclination, [deg] & \ang{28.19} & \ang{29.01} & \ang{43.87} \\
 & Final mass delivered, [kg] & 3892 & 3733 & 3257 \\ 
 & Propellant consumption, [kg]  & 3595 & 3754 & 4231 \\ 
 & $\Delta V$, [km/s] & 64.16  & 68.40  &  81.63 \\ 
\hline \hline
\end{tabular} 
\label{tab:maneuverscomp}
\end{table}
In Table \ref{tab:maneuverscomp} the summary for the interplanetary and rendezvous maneuver phases for the three real case missions is presented. $\Delta V$ differences and propellant consumption from Table \ref{tab:maneuverscomp} are mainly linked to the distance of the destination to reach and the final inclination to achieve, as well as the overall flight time. This means that, in order to keep the flight time reasonably low, the acceleration phase lasts a little longer for missions to the furthest dwarf planets. Also, part of the total acceleration duration is due to the required change of inclination: higher elevation from the ecliptic plane requires an increase in the out of plane thrust component direction, but this limits the possible acceleration, thereby there is a need to increase the thrust days. A trade-off between thrust vector direction and thrust days has been performed to minimize the fuel consumption. As far as the rendezvous is concerned, for the Eris mission the $\Delta V$ is about 10 km/s more demanding: this is because, in addition to the higher final coasting velocity, there is the fact that the velocity vector for that case has to be turned of about \ang{120} to match Eris velocity, while for the other two cases the angle between the initial and final velocities is about \ang{90}.
In Fig. \ref{fig:haumea28inc_vel_pos} the velocity evolution over time is reported for the Haumea mission. The overall evolution for the other mission is the same as the one presented in Fig. \ref{fig:haumea28inc_vel_pos}.  \par

For the Haumea mission, an analysis is performed on how the propellant consumption would be influenced if the payload mass is increased or decreased with respect to the nominal case of 1500 kg for missions of equal duration. The results are reported in Table \ref{tab:haumea_payloadmass_comp}.  \par

\begin{table} [h]
\centering
\caption{Comparison of propellant consumption for different payload masses for the Haumea mission. Here $ m_{p0} $ is the mass used as input for the trajectory design,  $m_i$ is the initial mass, $m_f$ is the final mass, $m_p$ is the effective payload mass and $m_{prop}$ is the propellant mass used. The final mass is the total mass which we have at destination, that is payload plus engine plus remaining fuel mass. The effective payload mass is considered to be the starting payload mass plus all the mass allocated for fuel that is not used.}
\begin{tabular}{ c c c c}
\hline \hline
 & Case 1 & Case 2 & Case 3 \\
\hline 
$m_{p0}$, [kg] & 1000 & 1500 & 2000 \\ 
$m_i$, [kg] & 6988 & 7488 & 8488 \\   
$m_f$, [kg] & 3579 & 3892 & 4216 \\   
$m_p$, [kg] & 1579 & 1892 & 2216 \\ 
$m_{prop}$, [kg] & 3409 & 3595 & 4272 \\ 
$m_f/m_i$ & 0.512 & 0.520 & 0.497 \\   
$m_p/m_i$ & 0.226 & 0.253 & 0.261  \\   
$m_p/m_{prop}$ & 0.463 & 0.526 & 0.519 \\ 
\hline \hline
\end{tabular} 
\label{tab:haumea_payloadmass_comp}

\end{table}
Let us analyze the results presented in Table \ref{tab:haumea_payloadmass_comp}. The first thing to note is that there is only a slight difference between the three final masses to initial masses ratio, $m_f/m_i$. One can expected the first case to be higher than the others, but it is important to consider that about 2000 kg of engine are fixed, so even though we have a launch mass lower than nominal, the propellant used is quite high anyway, as is seen in the payload to propellant masses ratio, $m_p/m_{prop}$. As expected, the $m_f/m_i$ ratio is the lowest for the third case, but that is because the initial payload mass is twice that of the first case. The $m_p/m_{prop}$ ratio, though, is very similar to that of the nominal case, which is case 2. From these considerations, an important observation on the engine can be derived: the 2 MW DFD engine is very effective for a payload that is over 1500 kg, but if the payload mass is decreased, it becomes excessive. In that case, it could be useful to consider a lower power DFD, so that is ideal for the time and mass constraints of the mission.  \par

\section{Mission Design for 125 AU destinations} \label{sec:125AU}
In this section we analyze two scenarios. The first one is a mission that has as its primary objective that of reaching 125 AU to study the magnetosphere of the Sun, but in its voyage it will also flyby Eris dwarf planet. The second one is a collection of missions in which the initial acceleration phase changes to understand its influence on the rest of the mission. Eris is chosen as the flyby destination because it is the most representative dwarf planet, due to its highly inclined and eccentric orbit and to the planet's physical characteristics, but any other TNO can be considered for the flyby. Also, the study on the acceleration phase can easily be applied to the first scenario.  \par

For these kinds of missions, the description of the different phases is omitted because they are the same as for the previous scenarios, with the only exception that here only the spiral departure phase and the acceleration and coasting phase are taken into account. The rendezvous phase is not required because there is no need to slow down the spacecraft at 125 AU: the objective is to travel through the edge of the heliosphere, and not to stop in a particular point. 
We choose to study this mission for a spacecraft with the same configuration as for the TNOs missions, so that  propellant masses used for different missions performed with the same engine become comparable. The launch mass is 7500 kg, powered by a 2 MW DFD, as for the TNOs missions. The spiral trajectory is the same as in previous scenarios: it has a duration of 79 days, a propellant usage of 556 kg and a $\Delta V$ of 7.55 km/s.
 \par

\subsection{125 AU destination with Eris flyby}
Results of calculations for the acceleration and coasting phases for the 125 AU mission with Eris flyby are given in Table \ref{tab:125eris}. As it is seen from Table \ref{tab:125eris}, in a little less than 9 years this mission aims to reach enormous distances from the Sun, visiting Eris and then fly towards interstellar space. This is possible because of the tremendous velocity reached at the end of the acceleration phase, which remains quite unaffected throughout the entire flight. \par

\subsection{125 AU destination with different acceleration phases}
The final step in this study is to understand how the acceleration period after the escape from Earth affects the missions to reach 125 AU. These calculations are performed both for a 2500 kg and 3000 kg initial dry masses, and results for the second case are presented in Table \ref{tab:125_acc}. The trend for the first case is exactly the same as in the second case, and for this reason the former is not reported here.  As it is clear from Table \ref{tab:125_acc}, a difference of 20 days during acceleration leads to half a year (about 6 months) less of flight time, and this corresponds to an additional propellant mass of less than 150 kg. Starting at 9 years, in both cases, the time of flight is reduced to about 7 years. This means that a difference of 80 days of thrust results in a 2 year gain. \par
\begin{figure}
\includegraphics[width=0.5\linewidth]{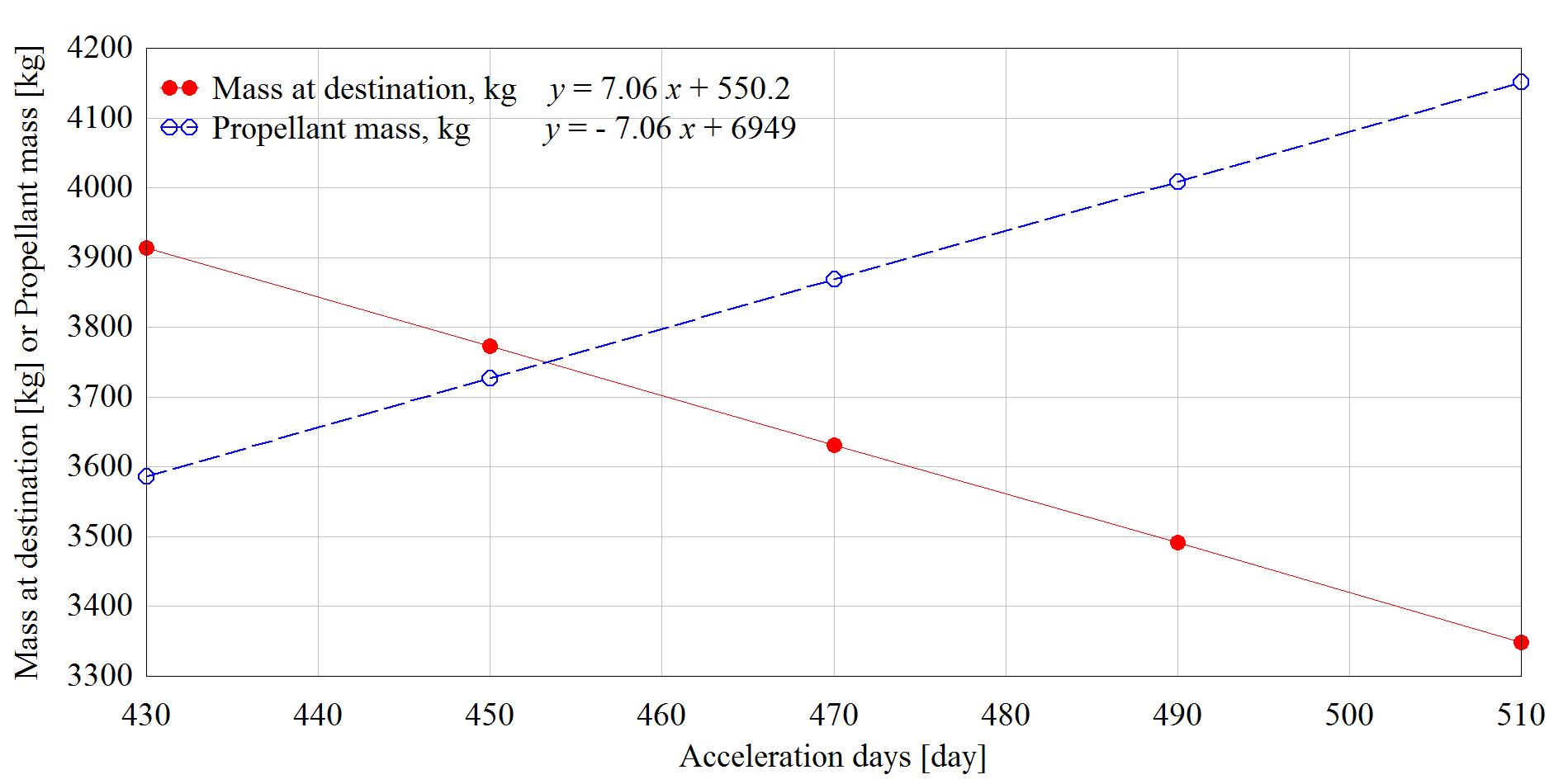}
\includegraphics[width=0.5\linewidth]{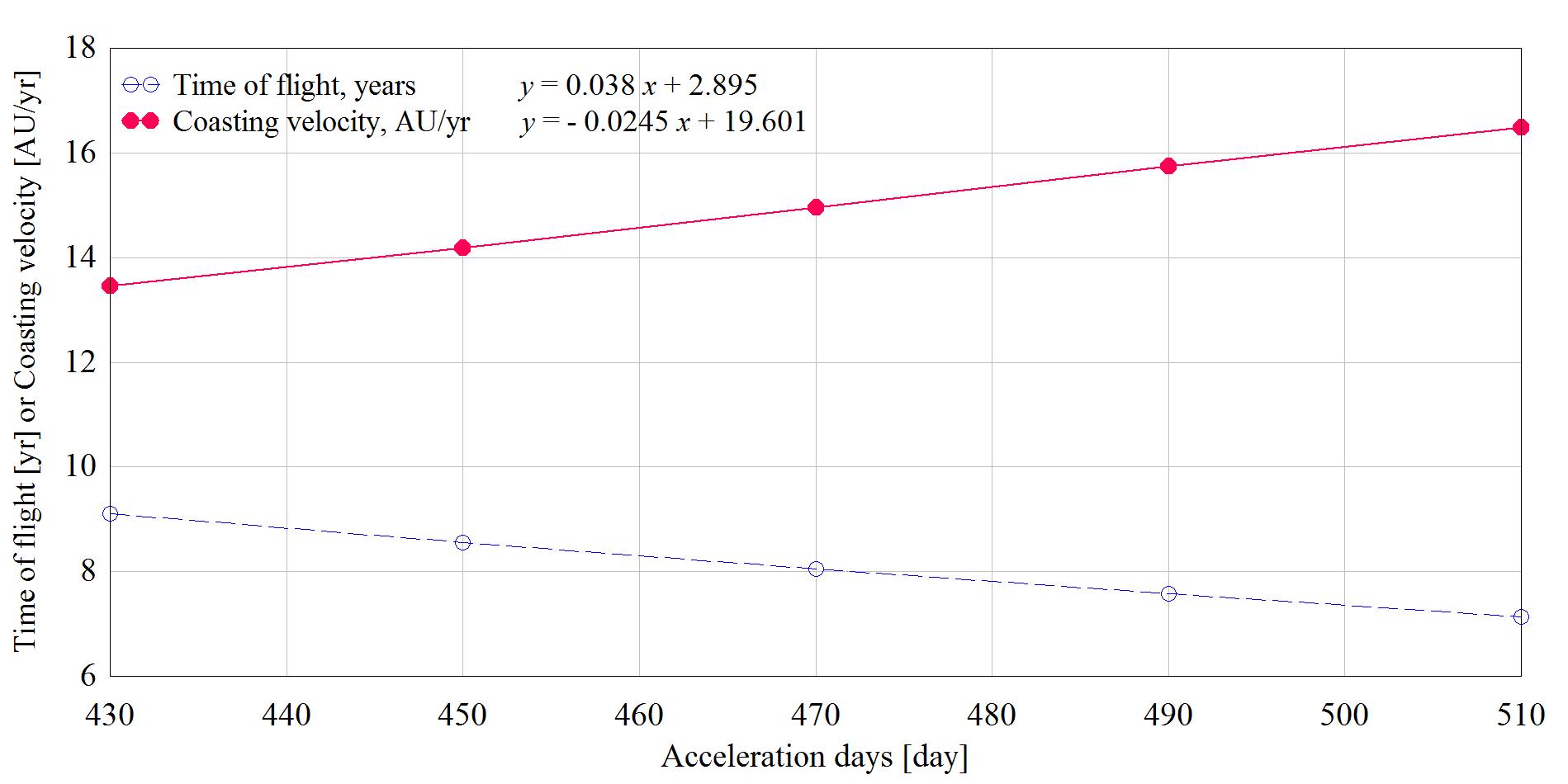}
\caption{Dependence of propellant and final masses (left panel) and time of flight and coasting velocity (right panel) on the acceleration days. The corresponding equations for linear interpolation of these dependencies are shown in the panels.}
\label{fig:acc_days_plot}
\end{figure}

One can use data from Table \ref{tab:125_acc} to get an analytic approximation  of these dependencies. The extrapolation of the propellant and destination masses, as well as time of flight and coasting velocity as a function of acceleration days leads to linear dependencies and are shown in Fig. \ref{fig:acc_days_plot}. In particular, the slopes of the lines for final and propellant masses have opposite sign: if the propellant mass increases, the final delivered mass decreases (left panel). On the other hand, the increase of coasting velocity leads to decrease of flight time (right panel).  \par

\begin{table} [h]
\centering
\caption{Acceleration and coasting phase for the 125 AU mission with Eris flyby. The initial Sun distance is the distance at which the thrust is turned off, while the distance at flyby is the distance of Eris from the Sun at the moment of flyby.}
\begin{tabular}{ r c }
\hline \hline
 & Acceleration phase \\ 
\hline 
Duration & 510 days \\
$\Delta V$ & 71.50 km/s \\
Fuel used & 3595 kg \\ 
Final velocity & 77.41 km/s (16.33 AU/yr)\\
\hline 
 & Coasting phase \\ 
\hline 
Duration & 7 years\\ 
Initial Sun distance & 11.34 AU \\ 
Distance at flyby & 80.23 AU\\ 
Final Sun distance & 125 AU \\  
Velocity at flyby & 76.55 km/s \\ 
\hline
 & Overall \\
\hline
Total mission duration,  & 8.67 years \\
Total $\Delta V$ & 79.05 km/s \\
Final mass delivered & 3350 kg \\
Propellant mass & 4150 kg \\
\hline \hline
\end{tabular} 
\label{tab:125eris}
\end{table}

\begin{table} [ht]
\caption{Overall results for a starting dry mass of 3000 kg. The propellant mass is the sum of propellant used in the acceleration phase and in the spiral phase. The latter is 556 kg for a period of 79 days. The time of flight is the elapsed time from the departure date from LEO. The coasting velocity is the speed of the spacecraft with respect to the Sun at the end of the acceleration phase and it will remain almost constant throughout the entire coasting phase.}
\centering
\begin{tabular}{c c c c c}
\hline \hline
Acceleration & Mass at  & Propellant & Time of  & Coasting  \\ 
days, [day]& destination, [kg] &  mass, [kg] & flight, [yr] & velocity, [AU/yr] \\
\hline 
430 & 3914 & 3586 & 9.10 & 13.46 \\  
450 & 3773 & 3727 & 8.56 & 14.19 \\  
470 & 3631 & 3869 & 8.05 & 14.95 \\  
490 & 3491 & 4009 & 7.58 & 15.75 \\  
510 & 3349 & 4151 & 7.14 & 16.58 \\ 
\hline \hline
\end{tabular} 
\label{tab:125_acc}
\end{table}

\section{Conclusions} \label{sec:conc}
In this work it is demonstrated that a spacecraft propelled by the DFD will enable an entirely new class of missions and will pave the way to a first practical approach to interstellar travel. The engine will be very versatile, and in one of its smaller configurations, such as that presented in this work, its thrust would be comparable to that of the most promising electromagnetic high power thrusters, but the specific impulse would be higher.
Our main objective is to present the possibilities to visit and even rendezvous with destinations such as TNOs and beyond using the DFD, given the time of flight and the launch mass. This kind of analysis is performed under multiple assumptions that, given the complexity of the mission and the distance of the target, brings us to some results that are very promising first estimations. It is worth noticing that our main goal here is to analyze and compare missions for a single realizable DFD rather than computing optimal solutions. For this reason, a single 2 MW DFD is chosen as engine, even though there is an optimal DFD and specific impulse for each mission. Tables \ref{tab:haumea_payloadmass_comp} and \ref{tab:125_acc} provide nice examples of trade-offs that are applicable to these studies. \par

So far, very few missions have been presented to study those worlds, and even fewer will be able to orbit the dwarf planets, because the deceleration required for a mission within 15 years would be impossible with the technology available at the moment, or at least it would be too demanding. The DFD, though, will be so powerful that it will be able of accelerate the spacecraft to extremely high velocities and then to decelerate it to the required speed, reducing of many years the time of flight. This makes its deceleration capability the real game changer.\par
  
Finally, one of the most important results found is related to the acceleration phases. A slight change in its duration will largely influence the total duration of the mission, as it is presented for the 125 AU mission, but it is also easily applicable to the TNOs missions. Furthermore, for the TNOs missions, acceleration and deceleration are tightly linked: for example, if 10 more days of acceleration are considered, they would result in roughly 10 more days of deceleration, and this increases the propellant consumption, but also decreases the total flight time. What we can infer from previous considerations is that such a propulsion system would enable the design team to tailor the trajectory to the needs of the mission, way more than any other propulsion system of the present days: this means that we can select the best possible solution according to mission drivers via a trade off between acceleration time and total flight time.\par

All the mission scenarios previously presented fit really well into the pool of missions already analyzed, both robotic and human \cite{cohen2017direct, ThomasPSS2017, razin2014direct, paluszek2014direct, genta2020dfd}. An engine like PPPL's DFD, if successfully developed, will be able to compete with the chemical thrusters for fast missions to inner planets and also will enable rapid missions to the outer planets at the border of the solar system, without the need of flybys. \par

\section*{Acknowledgments}
\noindent We are thankful to Prof. Cohen for his insightful discussions and to the team of Princeton Plasma Physics Laboratory for their hospitality and dedicated research, in particular Dr. Swanson. 

\bibliographystyle{elsarticle-num}
\biboptions{sort&compress}
{\setstretch{0.1}\bibliography{PA}}
\end{document}